# First-principles study of vibrational and dielectric properties of $\beta$-Si$_3$N$_4$

Yongqing Cai, Qingfeng Zeng, Litong Zhang, Laifei Cheng, Yongdong Xu

National Key Laboratory of Thermostructure Composite Materials,
Northwestern Polytechnical University, Xi'an, Shannxi 710072, PR China

ABSTRACT

First-principles calculations have been conducted to study the structural, vibrational and dielectric properties of $\beta$-Si$_3$N$_4$. Calculations of the zone-center optical-mode frequencies (including LO-TO splittings), Born effective charge tensors for each atom, dielectric constants, using density functional perturbation theory, are reported. The fully relaxed structural parameters are found to be in good agreement with experimental data. All optic modes are identified and agreement of theory with experiment is excellent. The static dielectric tensor is decomposed into contributions arising from individual infrared-active phonon modes. It is found that high-frequency modes mainly contribute to the lattice dielectric constant.

Silicon nitride (Si$_3$N$_4$) is an industrially important material due to its chemical, mechanical and electronic properties. Excellent thermo mechanical properties have seen this material used for engine parts, bearings, metal machining and other industrial applications.[1-3] Also, Si$_3$N$_4$ is considered to be a promising material for microelectronic applications because of its high dielectric constant and large electronic gap.[4]

Over the last three decades the lattice dynamical properties of Si$_3$N$_4$ have been investigated experimentally as well as theoretically by a number of groups. Brillouin zone center (BZC) phonon frequencies of Si$_3$N$_4$ have been measured by using Raman and infrared spectroscopies.[5, 6] However, identification of lattice modes in experimental studies is sometimes incomplete, partly because of the complexity of the structure and partly because of experimental errors. Earlier theoretical studies of dynamical properties of Si$_3$N$_4$ were done by adopting classical methods.[7, 8] Wendel[9] calculated the zone center frequencies through force field model derived using the Hessian biased technique from *ab initio* calculations. Although good agreement with experimental results for crystal structures, lattice expansion parameters, and the thermo-dynamical properties was obtained, the accuracy of frequency calculation was limited due to the types of interaction they took into account might not enough. Recently, first principles calculations of Raman active modes of β-Si$_3$N$_4$ have been reported.[10] The calculation was done by using a direct method where the phonon frequencies are calculated from Hellmann-Feynman forces generated by the small atomic displacement.

The first-principles study of phonon frequencies can be performed within a density functional perturbation theory (DFPT)[11] or direct method approach[12]. The advantage of the direct method is computationally straightforward and allows one to study phonons in the whole

Brillouin zone by perturbing the positions of the atoms slightly and calculating the reaction forces. However, one has to conduct a separate calculation using Berry's phase approach to obtain the LO-TO splitting of zone center optical modes. In contrast to the direct method of calculating the phonon frequencies, the linear response approach allows one to obtain the effective charges and dielectric tensors directly with no need to artificially increase the cell size in order to accommodate small values of the **q** vectors as in the direct method approach.

The purpose of this paper is to obtain in more detail the dynamical and dielectric properties of $\beta$-$Si_3N_4$ using the density functional perturbation theory. Our calculations start with the structural optimization and check that our relaxed structural parameters are consistent with the experimental work. The phonon frequencies and Born effective charge tensors are then computed from linear response techniques. Excellent agreement is found between the calculated frequencies and the measured spectra for both IR-active and Raman-active modes. The "missing" $A_g$ mode in Ref. 6 is identified. Finally, the dielectric properties and longitudinal-transverse (LO-TO) mode splittings are investigated, and our theoretical information is combined to predict the lattice contributions to the bulk dielectric tensor. We thus clarify the dependence of the dielectric response on orientation, and lattice dynamical properties and find that high-frequency modes mainly contribute to the lattice dielectric constant.

The calculations are carried using the CASTEP [13,14] code with Norm-conserving pseudopotentials.[15] The 2s and 2p semicore shells are included in the valence for N, and the 3s and 3p are included in the valence for Si. We employ the Ceperley-Alder[16] local density functional potential as parameterized by Vosko et al.[17] The kinetic energy cutoff for the plane waves is 770 ev, Brillouin zone integration is performed using a discrete $4 \times 4 \times 10$

Monkhorst-Pack[18] k-point sampling for a primitive cell. The first-principles investigation of vibrational and dielectric properties are conducted within a density functional perturbation theory (DFPT).[11] Phonon frequencies at the zone center, Born effective charge tensors, and dielectric tensors are computed as second-order derivatives of the total energy with respect to an external electric field or to atomic displacements. Technical details can be found in.[19, 20] To obtain LO/TO splitting characteristic for β-$Si_3N_4$ we introduced to the dynamical matrix the non-analytical term proposed by Pick.[21]

β-$Si_3N_4$ has a space group $P6_3/m$. The hexagonal unit cell contains two formula units (14 atoms). Figure I shows a ball-stick model of β-$Si_3N_4$. The silicon nitride structure can be described as a stacking of the idealized Si-N layers in an …**ABAB**… sequence. All Si atoms are equivalent (6h sites), but there are two inequivalent nitrogen sites: $N^{2c}$ at 2c sites and $N^{6h}$ at 6h sites. The $N^{2c}$ atoms are in a planar geometry with their three Si nearest neighbors, and the $N^{6h}$ atoms are in slightly puckered sites surrounded by three Si atoms, whereas Si atoms are at the center of slightly irregular tetrahedron bonded with one $N^{2c}$ atom and three $N^{6h}$ atoms.

The structural optimization was done by relaxing both the internal coordinates and the lattice constants by calculating the *ab initio* forces on the ions, within the Born-Oppenhaimer approximation, until the absolute values of the forces were converged to less than 0.01ev/Å. The relaxed lattice and fractional internal parameters of the equilibrium structure are listed in Table I and the initial configuration is taken from.[23] It can readily be seen that there is excellent agreement between our results and previous experiment. The volume are slightly underestimated, by 2%-3%, as is typical of LDA calculations. It should be noted that the accuracy in comparison with experimental lattice constants is much lower for two reasons: (a) DFT in the LDA is an approximation to the exact n-electron problem. (b) Temperature effects are neglected in the calculations; the calculations are performed for a fictitious classical system at zero temperature.

Since the primitive unit cell of β-Si$_3$N$_4$ structure has 14 atoms, there are a total of 42 modes of vibration. The irreducible representation at the center of the Brillouin zone is

$$\Gamma_{optic} = 4A_g(R) + 2E_{1g}(R) + 5E_{2g}(R) + 2A_u(IR) + 4E_{1u}(IR) + 3B_g + 2E_{2u}$$

$$\Gamma_{acoustic} = A_u + E_{1u}$$

Because of inversion symmetry, IR and Raman modes are mutually exclusive.

In Table II we compare calculated phonon frequencies at the Γ point with the measured Raman[5,6] and infrared[5] values. For comparison we cite also the phonon frequencies from theoretical calculations through direct methods[10] and classical model.[9] Our calculated Raman active mode frequencies present a rms absolute deviation of 8.6 cm$^{-1}$, and a rms relative deviation with of 1.9% with respect to the measurements of Ref. 6. This is an excellent agreement with both experiments and theory, and makes us very confident in the prediction of the frequencies for the IR active modes. It should be noted that the values of the vibrational properties of β-Si$_3$N$_4$ through the empirical model proposed by Wendel are just qualitative. Because of weak intensity and closeness to a relatively stronger E$_{2g}$ mode, the A$_g$ mode in the intermediate frequency band was misidentified[5,6] and is now known to be 457 cm$^{-1}$ through theoretical calculation.[10] Our calculation shows that this mode is at 456 cm$^{-1}$ which is in very good agreement with previous theoretical calculations.

The IR modes group into modes with displacements either in the x, y plane or along the z direction. The E$_{1u}$ mode has displacement pattern in the x, y plane, and the A$_u$ mode has displacements along z. For IR active mode, the root mean square relative deviation of our results with experimental data is 4.9%，with a slight overall tendency to underestimation. For some of the modes there are large differences with the reported experimental frequencies.[5] For example, our calculated frequency for the $E_{2u}$ mode is 886 cm$^{-1}$, whereas the experimental value is 985 cm$^{-1}$, the origin of these discrepancies is unclear. The LO/TO splittings for the infrared active mode are presented in Table Ⅲ. The LO/TO splittings for displacements parallel to the layers occur for the $E_{2u}$ mode and displacements parallel to c principal axis occur for the $A_u$ mode. The large LO/TO splittings of the three highest modes suggest they involve large effective charges and make large contributions to the static dielectric of β-Si$_3$N$_4$.

The Born effective charge tensor quantifies the macroscopic electric response of a crystal to the internal displacements of its atoms. Our results for the dynamical effective charges of β-Si$_3$N$_4$ are presented in Table Ⅳ for atoms at the 6h Wyckoff sites in the z=0.75 plane and two nitrogen atoms at 2c sites. The effective charges for the other atoms can be obtained from those shown in the table by symmetry considerations.

It is obvious that both Si and N atoms have lower effective charges than their formal charges, +4 and -3 for Si and N respectively, relating to the covalency of Si-N bonds. The Si atoms tend to show a more isotropic character than N atoms. As discussed in Sec. Ⅲ.A, the nitrogen atoms at the 2c sites are bonded to three nearest-neighbor silicon atoms in an planar configuration. One might then expect that the largest dynamical charge transfer would occur for motions of the N atoms in this plane. To check this, we computed the eigenvalues in the second column of table Ⅲ. Sure enough, the principal axis associated with the eigenvalue $Z_3^*$ = -1.76 of the smallest magnitude points normal to the plane of the neighbors. The other two principle axes lie in the plane of the neighbors. Not surprisingly in view of its tetrahedral coordination, the $Z^*$ tensor for atom Si is more isotropic, as indicated by the smaller spread of the eigenvalues in table.

Crystal symmetry makes the dielectric tensors be composed of some independent components. In hexagonal symmetry such as β-Si$_3$N$_4$, the calculated electronic ($\varepsilon_\infty$) and static ($\varepsilon_0$) dielectric tensors are diagonal and have two in dependent component $\varepsilon_\perp$ and $\varepsilon_\parallel$ along and perpendicular to the c axis, respectively. The electronic dielectric constants are calculated to be $\varepsilon_\perp^\infty = 4.19$ and $\varepsilon_\parallel^\infty = 4.26$. The reported value of $\varepsilon^\infty$ is 4.0 (Ref. 24), and our theoretical values tend to be overestimated in local density approximation (LDA) calculations due to the underestimation of the band gap.[25]

According to the generalized Lyddane-Sachs-Teller (LST) relation

$$\varepsilon_0 = \varepsilon_\infty \prod_m \frac{\omega_{LO,m}^2}{\omega_{TO,m}^2} \quad (1)$$

which is used separately for each polarization, the static dielectric constants $\varepsilon_0$ of both directions are $\varepsilon_\perp^0 = 8.79$ and $\varepsilon_\parallel^0 = 7.21$ respectively, showing nearly isotropic character. The dielectric properties of crystalline Si$_3$N$_4$ have not been studied very much experimentally. Experimental reports of the value of $\varepsilon_0$ for polycrystalline Si$_3$N$_4$ span a wide range from about 8.1 to 8.6 (Ref. 26), our orientationally average static dielectric is 8.26, falling comfortably in this range.

The static dielectric tensor $\varepsilon_0$ can also be separated into contributions arising from purely electronic screening $\varepsilon^\infty$ and IR-active phonon modes according to

$$\varepsilon_0 = \varepsilon_\infty + \varepsilon_\infty \sum_m \frac{\omega_{LO,m}^2 - \omega_{TO,m}^2}{\omega_{TO,m}^2} \quad (2)$$

which is different from the approximate "generalized Lyddane-Sachs-Teller" relation of Eq. (1). The differences are small, however. The static dielectric constants of both directions from Eq. (2) are $\varepsilon_\perp^0 = 7.67$ and $\varepsilon_\parallel^0 = 6.69$ compared to that of $\varepsilon_\perp^0 = 8.79$ and $\varepsilon_\parallel^0 = 7.21$ from Eq. (1).

For IR active modes the mode frequencies and contributions to the dielectric tensor are displayed in Table V. It can be seen that the high-frequency modes mainly contribute to the lattice dielectric constant. Although the mode contribution to the static dielectric constant is inversely proportional to the square of the mode frequency (Eq. (2)), low frequency modes $A_u$ at 378 cm$^{-1}$, $E_{2u}$ at 424 and 562 cm$^{-1}$ make relatively small contributions to the static dielectric tensor due to their smaller LO/TO splittings compared to higher frequency modes. When all the modes are summed over, we obtain the total lattice dielectric components of β-Si$_3$N$_4$ are $\varepsilon_\perp^{lat}=3.48$ and $\varepsilon_\parallel^{lat}=2.73$ compared to electronic contributions of $\varepsilon_\perp^\infty$=4.19 and $\varepsilon_\parallel^\infty$=4.26. We can conclude that β-Si$_3$N$_4$ has a lattice dielectric constant smaller than that of electronic contribution.

In conclusion, we have used density functional perturbation theory (DFPT) to investigate the vibrational and dielectric properties of β-Si$_3$N$_4$. Firstly, the structural parameters, including the internal coordinates, are relaxed, and excellent agreement is achieved experimental results. The vibrational modes at the center of the Brillouin zone have been evaluated and compared with experiment. All modes are identified and compared with experiments and previous theoretical calculations. The calculated zone-center phonon mode frequencies are in good agreement with infrared and Raman experiments. Finally, The Born effective charge tensors, and the dielectric permittivity tensors have also calculated. The mode contributions to the dielectric tensors have been obtained. We find that β-Si$_3$N$_4$ has a lattice dielectric constant smaller than that of electronic contribution.

The authors acknowledge support from the Natural Science Foundation of China (Contract No. 90405015), from the National Young Elitists Foundation (Contract No. 50425208) and from the Program for Changjiang Scholars and Innovative Research

Team in university (PCSIRT). We thank Northwestern Polytechnical University High performance Computing Center for allocation of computing time on their machines.

**REFERENCES**

[1] F. de Brito Mota, J. F. Justo, and A. Fazzio, Phys. Rev. B 58, 8323 (1998).

[2] F. L. Riley, J. Am. Ceram. Soc. 83, 245 (2000).

[3] R.N. Katz, Science 208, 84 (1980).

[4] M. J. Powell, B. C. Easton, and O. F. Hill, Appl. Phys. Lett. 38, 794 (1981).

[5] N. Wada, S. A. Solin, J. Wong, and S. Prochazka, J. Non-Cryst. Solids 43, 7 (1981)

[6] K. Honda, S. Yokoyama, and S. Tanaka, J. Appl. Phys. 85, 7380 (1999).

[7] C. M. Marian, M. Gastreich, and J.D. Gale, Phys. Rev. B 62, 3117 (2000).

[8] M. Gastreich, J. D. Gale, and C.M. Marian, Phys. Rev. B 68, 094110 (2003).

[9] J. A. Wendel and W. A. Goddard III, J. Chem. Phys. 97, 5048 (1992).

[10] J. Dong and O.F. Sankey, J. Appl. Phys. 87, 958 (2000).

[11] S. Baroni, S. de Gironcoli, A. Dal Corso, and P. Giannozzi, Rev. Mod. Phys. 73, 515 (2001).

[12] Ackland, G. J., Warren, M.C., Clark, S. J., J. Phys. Cond. Matt. 9, 7861 (1997).

[13] MD Segall, PJD Lindan, MJ Probert, CJ Pickard, PJ Hasnip, SJ Clark, and MC Payne, J. Phys.: Condens. Matter 14, 2717 (2002).

[14] Ackland, G.J.; Warren, M.C.; Clark, S. J., J. Phys.: Cond. Matt. 9, 7861 (1997).

[15] D. R. Hamann, M. Schlüter and C. Chiang, Phys. Rev. Lett. 43, 1494 (1979).

[16] D. M. Ceperley and B. J. Alder, Phys. Rev. Lett. 45, 566 (1980).

[17] S. H. Vosko, L. Wilk and M. Nusair, Can. J. Phys. 58, 1200 (1980).

[18] H. J. Monkhorst and J. D. Pack, Phys. Rev. B 13, 5188 (1976).

[19] X. Gonze, Phys. Rev. B 55, 10 337 (1997).

[20] X. Gonze and C. Lee, Phys. Rev. B 55, 10 355 (1997).

[21] R. Pick, M. H. Kohen, and R. M. Martin, Phys. Rev. B 1, 910 (1970).

[22] M. Billy, J.-C. Labbe, A. Selvaraj, and G. Roult, Mater. Res. Bull. 18, 921 (1983).

[23] O. Borgen and H. M. Seip, Acta Chem. Scand. 15, 1789 (1961).

[24] S. K. Andersson, et al., Optical Materials 10, 85 (1998).

[25] Ph. Ghosez, J.-P. Michenaud, and X. Gonze, Phys. Rev. B 58, 6224 (1998).

[26] M. K. Park, et al., Key Engineering Materials, 287, 247 (2005).

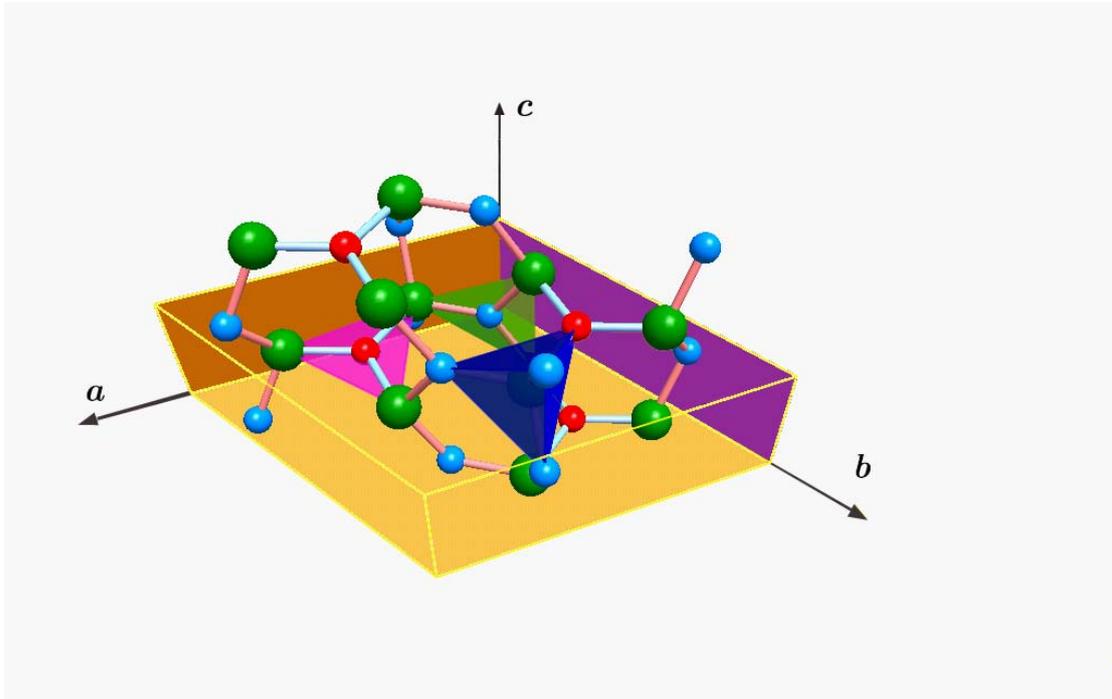

FIG. Ⅰ. A ball-stick model of β-Si$_3$N$_4$. The green, red and blue spheres represent Si atoms at 6h sites, N atoms at 2c sites, N atoms at 6h sites, respectively. N atoms are within the same plane of their three nearest neighbors, whereas Si atoms lie at the center of a slightly irregular tetrahedron.

TABLE I. Optimized geometry of β-Si$_3$N$_4$ (P63/m)

| Parameter | Experiment[a] | Experiment[b] | This work |
|---|---|---|---|
| V (Å$^3$) | 147.016 | 145.881 | 142.700 |
| c (Å) | 7.6272 | 7.6070 | 7.5572 |
| a (Å) | 2.9182 | 2.9110 | 2.8852 |
| x Si(6h) | 0.1842 | 0.1740 | 0.1742 |
| y Si(6h) | 0.7704 | 0.7660 | 0.7678 |
| z Si(6h) | 0.2500 | 0.2500 | 0.2500 |
| x N(2c) | 0.3333 | 0.3333 | 0.3333 |
| y N(2c) | 0.6667 | 0.6667 | 0.6667 |
| z N(2c) | 0.2500 | 0.2500 | 0.2500 |
| x N(6h) | 0.3399 | 0.3210 | 0.3302 |
| y N(6h) | 0.0357 | 0.0250 | 0.0299 |
| z N(6h) | 0.2500 | 0.2500 | 0.2500 |

[a]Reference 22.
[b]Reference 23.

Table II. Zone centre optic phonon frequencies of $\beta$-$Si_3N_4$ (cm$^{-1}$)

| Mode | Expt.[a] (144 $A_g$) | Expt.[b] (145 $A_g$) | Theor.[c] | Theor.[d] | This work |
|---|---|---|---|---|---|
| Raman | | | | | |
| $E_{2g}$ | 186 | 185 | 183 | 190 | 181 |
| $A_g$ | 210 | 208 | 201 | 236 | 200 |
| $E_{1g}$ | 229 | 230 | 228 | 215 | 225 |
| $E_{2g}$ | 451 | 452 | 444 | 518 | 444 |
| $A_g$ | … | … | 457 | 289 | 456 |
| $E_{2g}$ | 619 | 620 | 603 | 592 | 610 |
| $A_g$ | 732 | 733 | 715 | 539 | 725 |
| $E_{1g}$ | 865 | 866 | 836 | 975 | 859 |
| $E_{1g}$ | 928 | 930 | 897 | 1017 | 921 |
| $A_g$ | 939 | 940 | 908 | 547 | 930 |
| $E_{2g}$ | 1047 | 1048 | 1012 | 1067 | 1035 |
| Infrared | | | | | |
| $A_u$ | 380 | | | | 378 |
| $E_{2u}$ | 447 | | | | 424 |
| $E_{2u}$ | 580 | | | | 562 |
| $A_u$ | 910 | | | | 848 |
| $E_{2u}$ | 985 | | | | 886 |
| $E_{2u}$ | 1040 | | | | 1021 |

[a]Reference 5.  [c]Reference 10.
[b]Reference 6.  [d]Reference 9.

Table III. LO/TO splittings for infrared active modes of β-Si$_3$N$_4$. The first column(E=0) is for no electric field, the second is for the field lying in the plane ($E \parallel a-b$), and the third is for $E \parallel c$. $\Delta\omega^2 = \omega_{LO}^2 - \omega_{TO}^2$.

| Mode | Phonon frequency (cm$^{-1}$) | | | $\sqrt{\Delta\omega^2}$ (cm$^{-1}$) |
| --- | --- | --- | --- | --- |
| | E=0 | E $\parallel$ a-b | E $\parallel$ c | |
| A$_u$ | 378 | 378 | 395 | 115 |
| E$_{2u}$ | 424 | 453 | 424 | 159 |
| E$_{2u}$ | 562 | 584 | 562 | 159 |
| A$_u$ | 848 | 848 | 1056 | 629 |
| E$_{2u}$ | 886 | 1026 | 886 | 517 |
| E$_{2u}$ | 1021 | 1150 | 1021 | 529 |

Table IV. The Born effective charges of β-Si$_3$N$_4$. $Z_j^*$ (j=1,2,3) is the jth eigenvalue of the symmetric part of the $Z^*$ tensor. $\overline{Z}$ is average of eigenvalues. Only atoms at the 6h Wyckoff sites in the z=0.75 plane are presented.

|  | $N_1^{2c}$ | $N_2^{2c}$ | $N_1^{6h}$ | $N_2^{6h}$ | $N_3^{6h}$ | $Si_1^{6h}$ | $Si_2^{6h}$ | $Si_3^{6h}$ |
|---|---|---|---|---|---|---|---|---|
| $Z_1^*$ | -2.95 | -2.95 | -1.61 | -3.15 | -1.61 | 3.15 | 3.58 | 3.15 |
| $Z_2^*$ | -2.95 | -2.95 | -3.15 | -1.61 | -3.15 | 3.58 | 3.15 | 3.58 |
| $Z_3^*$ | -1.76 | -1.76 | -2.82 | -2.82 | -2.82 | 3.41 | 3.41 | 3.41 |
| $\overline{Z}$ | -2.55 | -2.55 | -2.53 | -2.53 | -2.53 | 3.38 | 3.38 | 3.38 |

Table V. IR active mode frequency, contribution to the component of dielectric tensor for each IR-active mode. The component of lattice dielectric constant parallel (perpendicular) to c-axis is denoted $\varepsilon_\parallel^{lat}$ ($\varepsilon_\perp^{lat}$).

| Mode | $\omega_{TO}$ (cm$^{-1}$) | $\omega_{LO}$ (cm$^{-1}$) | $\sqrt{\Delta\omega^2}$ (cm$^{-1}$) | $\varepsilon_\perp^{lat}$ | $\varepsilon_\parallel^{lat}$ |
|---|---|---|---|---|---|
| A$_u$ | 378 | 395 | 115 | 0 | 0.39 |
| E$_{2u}$ | 424 | 453 | 159 | 0.59 | 0 |
| E$_{2u}$ | 562 | 584 | 159 | 0.34 | 0 |
| A$_u$ | 848 | 1056 | 629 | 0 | 2.34 |
| E$_{2u}$ | 886 | 1026 | 517 | 1.43 | 0 |
| E$_{2u}$ | 1021 | 1150 | 529 | 1.12 | 0 |